\documentclass{pasj00}
\draft

\begin{document}
\SetRunningHead{M. Sato and H. Asada}
{Mutual Transits by Extrasolar Planet-Companion} 
\Received{}
\Accepted{}

\title{Effects of Mutual Transits by Extrasolar 
Planet-Companion Systems on Light Curves} 

\author{Masanao \textsc{Sato}, 
and  
Hideki \textsc{Asada}} %
\affil{Faculty of Science and Technology, Hirosaki University, 
Hirosaki, Aomori 036-8561}


%


\maketitle

\begin{abstract}
We consider the effects of mutual transits by 
extrasolar planet-companion systems 
(in a true binary or a planet-satellite system) 
on light curves. 
We show that induced changes in light curves 
depend strongly on a ratio between a 
planet-companion's orbital velocity around their host star 
and a planet-companion's spin speed 
around their common center of mass.  
In both the slow and fast spin cases 
(corresponding to long and short distances between them, 
respectively), 
a certain asymmetry appears in light curves. 
We show that, 
especially in the case of short distances, 
occultation of one faint object by the other, 
while the transit of the planet-companion system 
occurs in front of its parent star, causes an apparent increase 
in light curves and characteristic fluctuations appear 
as important evidence of mutual transits. 
We show also that extrasolar mutual transits provide 
a complementary method of measuring the radii 
of two transiting objects, 
their separation and mass, 
and consequently identifying them as a true binary, 
planet-satellite system or others. 
Monitoring $10^5$ stars for three years with {\it Kepler} 
may lead to a discovery of a second Earth-Moon-like system 
if the fraction of such systems for an averaged star is 
larger than 0.05, 
or 
it may put upper limits on the fraction as $f < 0.05$. 
\end{abstract}

\section{Introduction}
It is of general interest to 
discover a second Earth-Moon system. 
Detections of extrasolar planet-satellite or binary planet 
systems will bring important information to 
planet (and satellite) formation theory 
(e.g., Jewitt and Sheppard 2005, 
Canup and Ward 2006, 
Jewitt and Haghighipour 2007). 

It is not clear whether 
the IAU definition for planets in the solar system 
can be applied to extrasolar planets as it is. 
The IAU definition in 2006 is as follows. 
A planet is a celestial body that 
(a) is in orbit around the Sun, 
(b) has sufficient mass so that it assumes a hydrostatic equilibrium 
(nearly round) shape, and 
(c) has cleared the neighborhood around its orbit.

We call a gravitationally bound system of two extrasolar 
planet-size objects simply as extrasolar {\it binary planets}. 
They constitute a true binary 
if the following conditions are satisfied instead of (c) 
in addition to the two criteria (a) with replacing the Sun 
by a host star and (b). 
(c1) Their {\it total} mass is dominant 
in the neighborhood around their orbits. 
(c2) Their common center of mass is 
{\it above} their surfaces. 
If it is below a surface of one object, 
one may call them an extrasolar planet-satellite. 

There have been theoretical works on the existence of planets 
with satellites. 
The solar system's outer gaseous planets have multiple 
satellites, each of which notably has a similar fraction 
($\sim 10^{-4}$) of their respective planet's mass. 
For instance, Canup and Ward (2006) found that 
the mass fraction is regulated to $\sim 10^{-4}$ 
by a balance between two competing processes 
of the material inflow to the satellites and 
the satellite loss through orbital decay driven by the gas.  
They suggested that similar processes could limit 
the largest satellite of extrasolar giant planets. 
Such theoretical predictions await future observational tests. 
There still remains a possibility of detecting Jupiter-size binary 
planets with comparable masses. 
Furthermore, we should note that their model does not hold 
for solid planets. 
It may be possible to detect binary solid planets 
(perhaps Earth-size ones). 
Therefore, future detection of extrasolar planet-companion systems or 
a larger mass fraction ($> 10^{-4}$) of satellites 
around gaseous exoplanets will 
give definite information on the planet and satellite 
formation theory. 
In any case, unexpected findings will open the possibility 
of new configurations such as binary planets.

Recent direct imaging of a planetary mass $\sim 8 M_J$ 
with an apparent separation of 330 AU from the parent star 
(Lafreni{\`e}re et al. 2008) 
indicates the likely existence of long-period exoplanets 
($> 1000$ yr). 
In this paper, we consider such exoplanets as well as close ones.  

Since the first detection of a transiting extrasolar planet 
(Charbonneau et al. 2000), photometric techniques have been 
successful (e.g., Deming, Seager, Richardson, Harrington 2005 
for probing atmosphere, 
Ohta et al. 2005, Winn et al. 2005, Gaudi \& Winn 2007, 
Narita et al. 2007, 2008 for measuring stellar spins). 
In addition to 
{\it COROT}\footnote{http://www.esa.int/SPECIALS/COROT/}, 
{\it Kepler}\footnote{http://kepler.nasa.gov/} 
has been very recently launched. 
It will monitor about $10^5$ stars with 
expected 10 ppm ($=10^{-5}$) photometric differential sensitivity. 
This enables the detection of a Moon-size object. 
 
Sartoretti and Schneider (1999) first 
suggested a photometric detection of extrasolar satellites. 
Cabrera and Schneider (2007) developed a method 
based on the imaging of a planet-companion as an unresolved 
system (but resolved from its host star) 
by using planet-companion mutual transits and mutual shadows. 
As an alternative method, 
timing offsets for a single eclipse have been investigated 
for eclipsing binary stars 
as a perturbation of transiting planets around the center of mass 
in the presence of the third body 
(Deeg et al. 1998, 2000, Doyle et al. 2000). 
It has been recently extended toward detecting ``exomoons''  
(Szab{\'o}, Szatm{\'a}ry, Div{\'e}ki, Simon 2006, 
Simon, Szatm{\'a}ry, Szab{\'o}, 2007, 
Kipping 2009a, 2009b). 
The purpose of the present paper is to investigate 
effects of mutual transits by extrasolar planet-companion 
systems on light curves, 
especially how the effects depend on their spin velocity 
relative to their orbital one around their parent star. 
Furthermore, we shall discuss extrasolar mutual transits 
as a complementary method of measuring the system's parameters  
such as a planet-companion's separation 
and thereby of identifying them as a true binary, 
planet-satellite system or others. 

Our treatment is applicable both to a true binary and 
to a planet-satellite system. 
Our method has analogies in classical ones for eclipsing binaries 
(e.g., Binnendijk 1960, Aitken 1964). 
A major difference is that 
occultation of one faint object by the other 
transiting a parent star causes an apparent {\it increase}  
in light curves, 
whereas eclipsing binaries make a decrease. 
What is more important is that, in both cases where one faint object 
transits the other and vice versa, 
changes are made in the light curves due to mutual transits 
even if no light emissions come from the faint objects. 
In a single transit, on the other hand, thermal emissions from 
a transiting object at lower temperature make a difference 
in light curves during the secondary eclipse, when the object 
moves behind a parent star as observed for instance for HD209458b 
(Deming et al. 2005).

Let us briefly mention transits/occultations in the solar system. 
It is possible that the Moon or another celestial body 
occult multiple celestial bodies at the same time. 
Such mutual occultations are extremely rare and 
can be seen only from a small part of the world. 
The last event was on 23rd of April, 1998, 
when the Moon occulted the Venus and Jupiter simultaneously 
for observers on Ascension Island. 
Such an event is extremely rare because 
it is controlled by three different orbital periods 
of the Moon, Venus and Jupiter 
and hence the probability of the alignment 
of the three objects is very low. 

In the case of planet-companion systems, on the other hand, 
orbital periods around a host star are common for the two objects. 
Therefore, the number of time scales controling 
extrasolar mutual transits are {\it two}; 
the orbital period around the host star and 
the planet-companion's spin period. 
As a result, extrasolar mutual transits  
(with one planet occasionally transiting or occulting the other) 
across a parent star, 
can occur more frequently 
than mutual occultations in our solar system. 

This paper is organized as follows. 
In section 2, we consider mutual transits of extrasolar 
planet-companion systems  
in orbit around their host star. 
Event rates and possible bounds are also discussed. 
Section 3 is devoted to the conclusion.

\section{Mutual Transits of Extrasolar Planet-companion Systems} 
\subsection{Approximations and notation}
For simplicity, we assume that the orbital plane of a 
planet-companion system 
orbiting around its common center of mass (COM) is the same as 
that of the COM in orbit around the host star with radius $R$. 
This co-planar assumption is reasonable 
because it seems that planets are born from fragmentations 
of a single proto-stellar disk 
and thus their spins and orbital angular momentum are 
nearly parallel to the spin axis of the disk. 

Inclination angles of the orbital plane with respect to 
our line of sight are chosen as 90 degrees, 
because transiting planets can be observed 
only for (nearly) the edge-on case.  
In order to show clearly our idea, 
the binaries are in circular motion 
as an approximation by (c1). 
It is straightforward to extend to elliptic motions. 

For investigating transits, we need 
the transverse position $x$ and velocity $v$. 
We denote those of 
the COM for planet-companion systems  
as $x_{CM}$ and $v_{CM}$, respectively, 
where the origin of $x$ is chosen as the center of the star.  
The position and velocity of each planet 
with mass $m_1$ and $m_2$  
in the binary system with separation $a$ 
are denoted as 
$x_i$ and $v_i$ ($i=1, 2$), respectively. 
We express the position of each planet as 
\begin{eqnarray}
x_1 
&=& 
x_{CM}+a_1 \cos\omega (t-t_0) , 
\label{x1}
\\
x_2 
&=& 
x_{CM}-a_2 \cos\omega (t-t_0) ,
\label{x2}
\end{eqnarray}
where 
the orbital radius of each planet around their COM 
is denoted by $a_i$,  
the angular velocity
of the binary motion is denoted by $\omega$, 
and 
$t_0$ means the time when the binary separation 
becomes perpendicular to our line of sight 
(See Table 1 for a list of parameters and their definition). 
For simplicity, we shall set $t_0 = 0$ below. 

One can approximate $v_{CM}$ as being constant during the transit, 
because the duration is much shorter than the orbital period 
for the binary around the host star.

\subsection{Transits in light curves}
The decrease in the apparent luminosity due to mutual transits 
is expressed as 
\begin{equation}
L = \frac{S-\Delta S}{S} , 
\label{L}
\end{equation}
where 
$S=\pi R^2$, 
$S_1=\pi r_1^2$,
$S_2=\pi r_2^2$,
$\Delta S=S_1+S_2-S_{12}$. 
Here, $r_1$ and $r_2$ denote the radii of the planets 1 and 2, 
and 
$S_{12}$ denotes the area of 
the apparent overlap between them, 
which is seen from the observer. 
Without loss of generality, we assume $r_1 \geq r_2$.

\subsection{Effects on light curves}
We investigate light curves 
by mutual transits due to planet-companion systems.  
The time derivative of Eq. ($\ref{x2}$) becomes 
$v_2 = v_{CM} + a_2 \omega\sin\omega (t-t_0)$ . 
Hence, the apparent retrograde motion is observed 
if $v_{CM} < a_2 \omega$, 
which we call the fast planet-companion's spin. 
If $v_{CM} > a_2 \omega$, we call it slow spin.  
The Earth-Moon and Jupiter-Ganymede systems represent 
slow and marginal cases, respectively. 

\begin{figure}
    \FigureFile(130mm,120mm){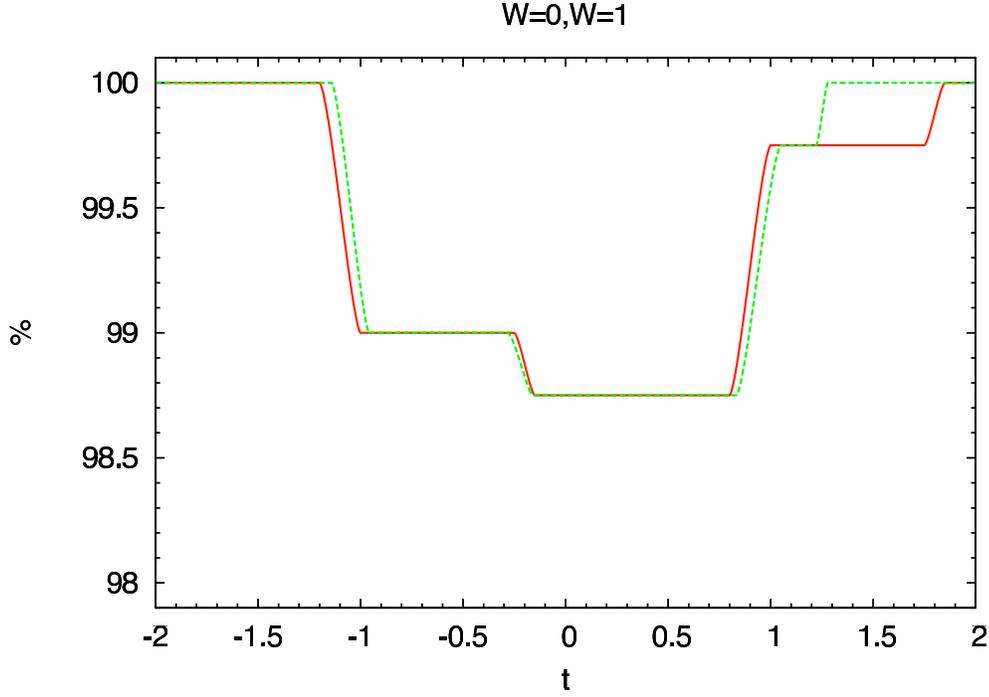}
\caption{
Light curves: 
Solid red one denotes 
the zero binary's spin limit as a reference 
($W \equiv a\omega/v_{CM} = 0$). 
Dashed green one is a slow spin case (large separation) for $W=1$. 
The vertical axis denotes the apparent luminosity 
(in percents). 
The horizontal one is time in units of 
the half crossing time of the star by 
the COM of the binary, defined as $R/v_{CM}$. 
For simplicity, we assume the binary with a common mass density, 
a radius ratio as 
$R: r_1: r_2 =20: 2: 1$, 
and $a/R=0.9$. 
}
\label{lightcurve-1}
\end{figure}

Figure $\ref{lightcurve-1}$ shows light curves by mutual transits 
for two cases. 
One is the zero spin limit of $\omega \to 0$ 
as a reference. 
In this case, motion of the two objects 
is nothing but a translation. 
Because of the time lag between the first and second transits, 
a certain plateau appears 
in light curves. 
The other is a slow spin case as $W=1$, where
$W$ denotes the dimensionless spin ratio defined as 
$a\omega/v_{CM}$. 
Its basic feature is the same as that for zero spin case, 
except for certain changes that are due to the relative motion 
between the planets. 
In the slow case, averaged inclination of each slope 
in the light curve is dependent on time, 
especially at the start and end of the mutual transit. 
Here we assume planet-companion systems with a common mass density, 
a radius ratio as 
$R: r_1: r_2 =20: 2: 1$, and $a/R=0.9$. 

Fast spin cases are shown by 
Figs. $\ref{lightcurve-3}$ and $\ref{lightcurve-4}$ 
($W=3$ and $6$, respectively), 
where the apparent retrograde motion 
produces characteristic fluctuations. 
Here we assume the same configuration as that 
in Fig. $\ref{lightcurve-1}$ except for shorter distance 
from the host star. 
These figures show also the transverse positions of planets with time, 
which would help us to understand the chronological changes 
in the light curves. 
In particular, 
it can be understood that such characteristic patterns appear 
only when two faint objects are in front of the star and 
one of them transits (or occults) the other. 

\subsection{Parameter determinations through mutual transits} 
\begin{figure}
    \FigureFile(130mm,120mm){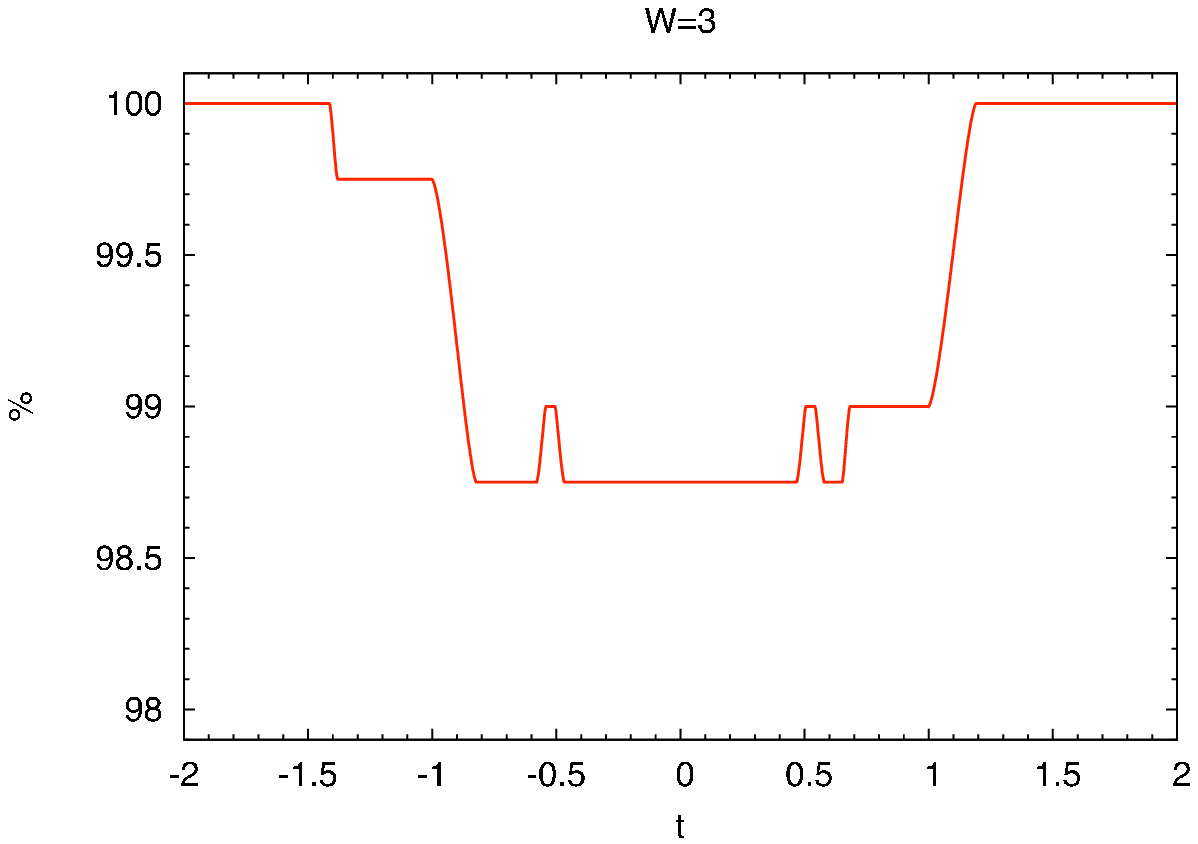}
\\
    \FigureFile(130mm,120mm){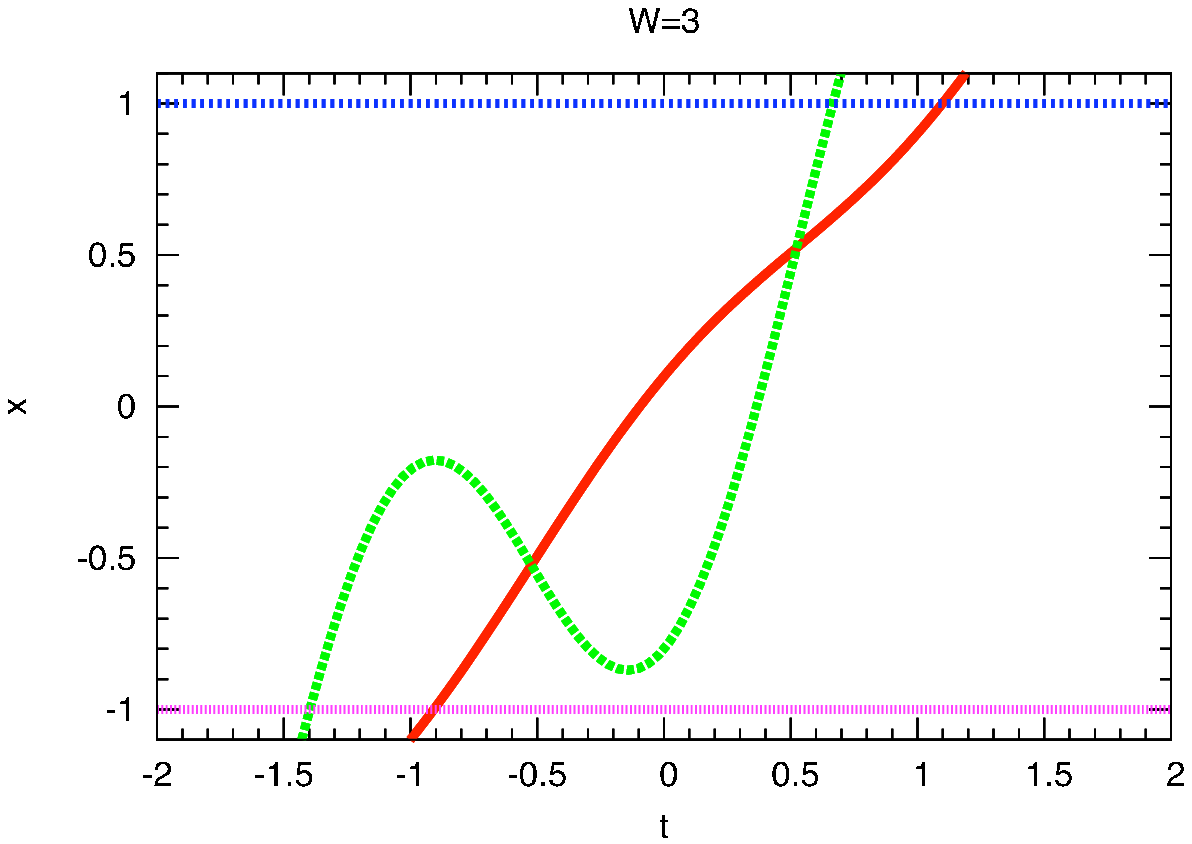}
\caption{
Top panel: a light curve for a fast spin case (small separation). 
The radius and mass ratio are the same as those 
in Fig. $\ref{lightcurve-1}$. 
We assume $W=3$. 
Brightness fluctuations appear with the width of 
$T_{top}=0.033$ and $T_{bottom}=0.1$. 
These values satisfy Eqs. ($\ref{Ttop}$) and ($\ref{Tbottom}$). 
Bottom panel: the motion of each body 
in the direction of $x$ normalized by $R$ 
(solid red for the primary and dotted green for the secondary). 
When one faint object transits or occults the other 
in front of the host star, 
mutual transits occur 
and a ``hill'' appears in the light curve. 
}
\label{lightcurve-3}
\end{figure}

\begin{figure}
    \FigureFile(130mm,120mm){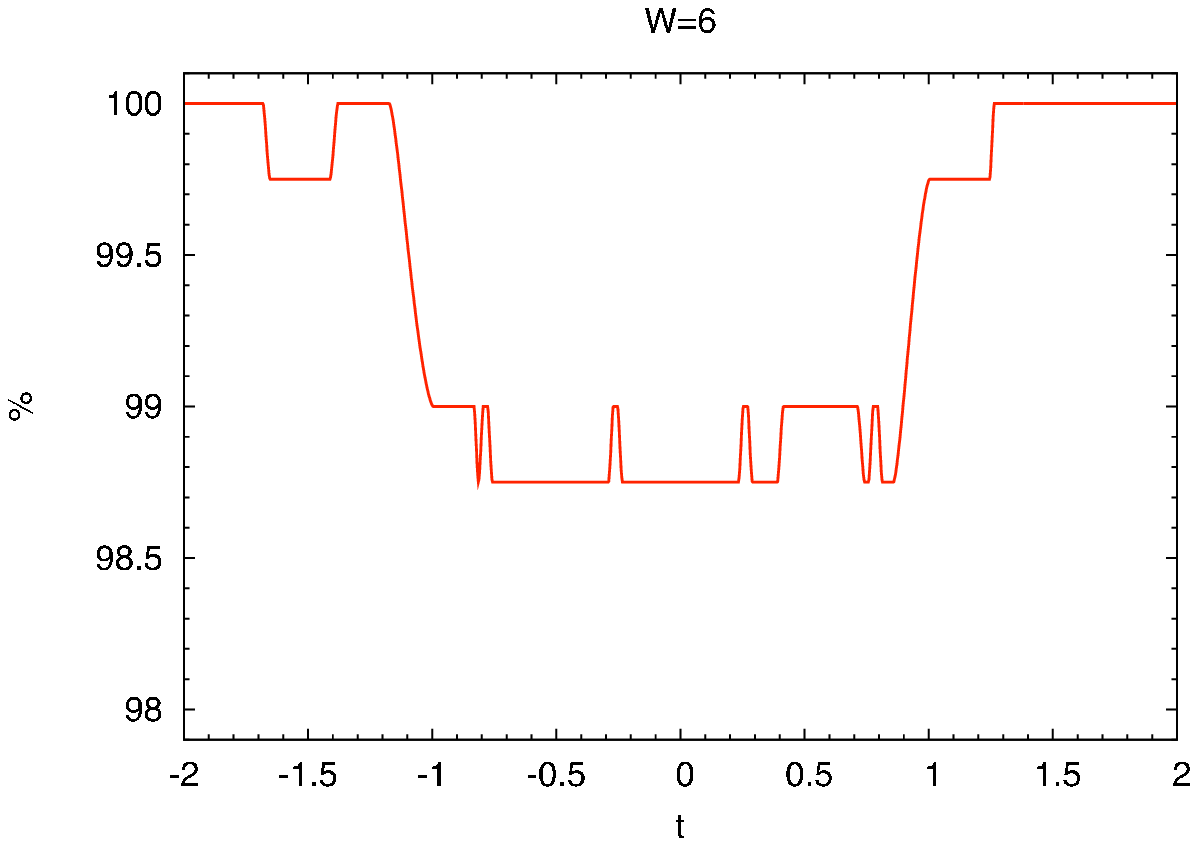}
\\
    \FigureFile(130mm,120mm){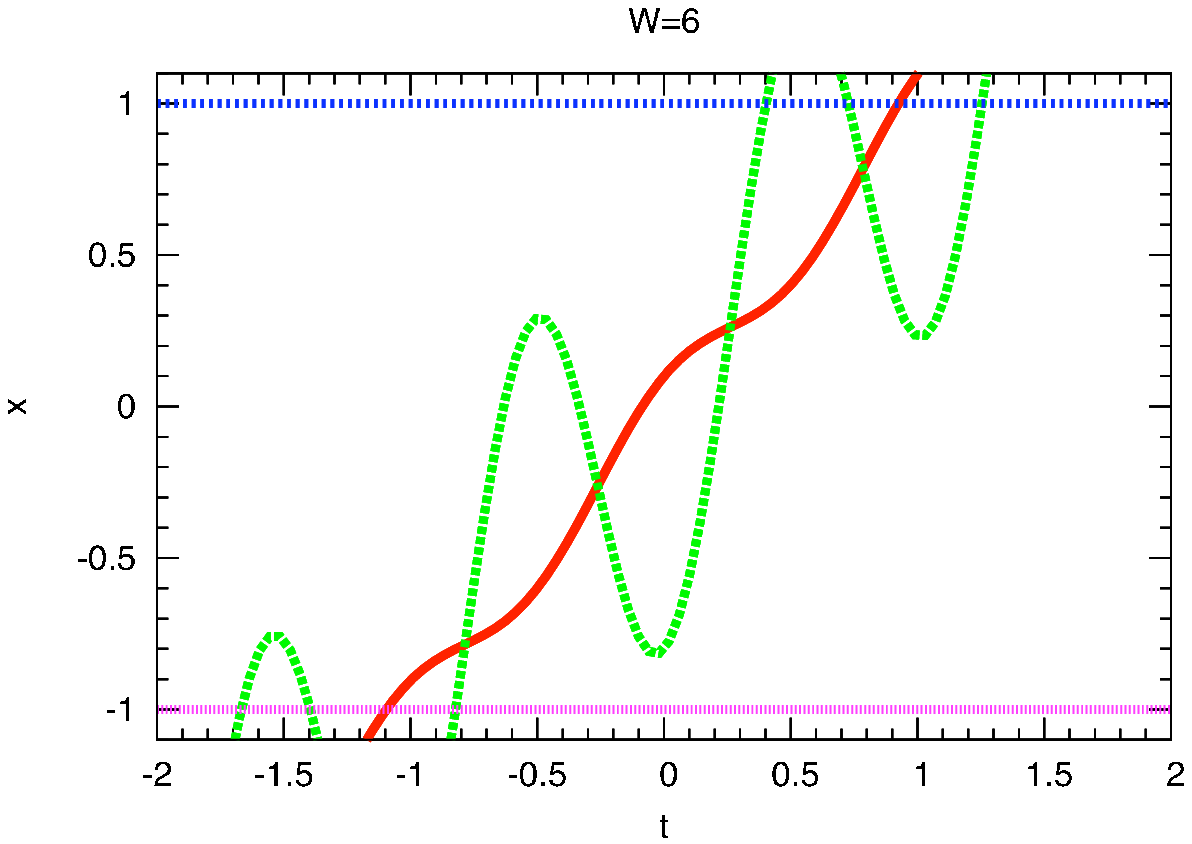}
\caption{
Top panel: a light curve for a  
faster spin case (smaller separation). 
Bottom panel: the motion of each body. 
The parameters are the same as those in Fig. $\ref{lightcurve-1}$, 
except for $W=6$. 
Comparing with Figure $\ref{lightcurve-3}$,  
the number of ``hills'' increases, 
and the shape of the light curve 
becomes more complicated especially at the bottom. 
A plateau around $t=0.5$ is due to a single transit 
of one faint object since the other has passed across a host star. 
}
\label{lightcurve-4}
\end{figure}

In all the above cases, the amount of decrease in light curves 
or the magnitude of fluctuations gives the ratios among the radii of 
the star and two faint objects $(R, r_1, r_2)$.  
Through behaviors of apparent light curves 
in both slow and fast cases, 
$a\omega$ (as its ratio to $v_{CM}$) can be obtained 
as shown by Figs. $\ref{lightcurve-1}$-$\ref{lightcurve-4}$. 
Here, $v_{CM}$ is determined as $v_{CM} = 2R/T_E$ 
by measuring the duration of the whole transit time $T_E$ 
because of $R \gg r_1, r_2$,  
if the stellar radius $R$ (and mass $m_S$) 
are known for instance by its spectral type.
Therefore, $a\omega$ is determined separately. 
The planet-companion's spin velocity $a\omega$ determines the gravity 
between the objects. 

The spin period $P$ (and thus $\omega$) can be determined, 
especially for the fast rotation case that produces multiple 
``hills'', 
because an interval between neighboring ``hills'' is nothing but 
a half of the binary period. 
As a result, the binary separation $a$ is obtained separately. 
Hence one can determine the total mass of the binary 
as $Gm_{tot}=\omega^2 a^3$ from Kepler's third law, 
where $G$ denotes the gravitational constant. 

If we assume also that the mass density is common 
for two objects constituting the binary 
(this may be reasonable especially for similar size objects 
as $r_1 \sim r_2$), 
each mass is determined as $m_1 = r_1^3 (r_1^3 + r_2^3)^{-1} m_{tot}$ 
and $m_2 = r_2^3 (r_1^3 + r_2^3)^{-1} m_{tot}$, respectively. 
Therefore, the orbital radius of each body around the COM 
is obtained as $a_1 = r_2^3 (r_1^3 + r_2^3)^{-1} a$ 
and $a_2 = r_1^3 (r_1^3 + r_2^3)^{-1} a$, respectively. 
At this point, importantly, the two objects can be identified as 
a true binary or planet-satellite system. 

In a slow spin case, on the other hand, 
the apparent separation $a_{\perp}$ (normal to our line of sight) 
is determined as $a_{\perp} = T_{12} v_{CM}$ from measuring the time lag
$T_{12}$ between the first and second 
transits because $v_{CM}$ is known above. 

\begin{figure}
    \FigureFile(130mm,120mm){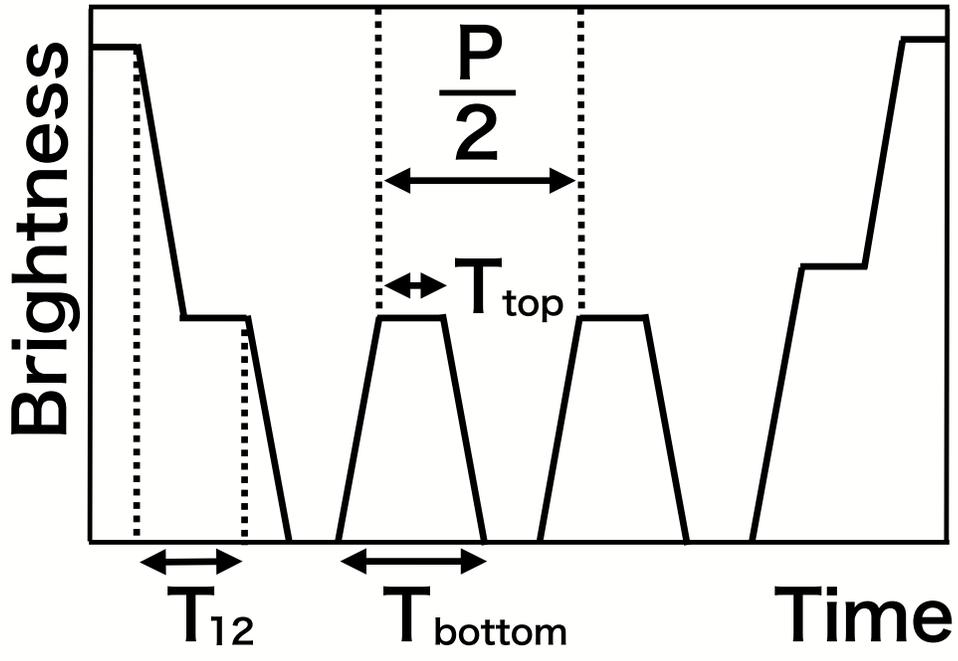}
\caption{
Schematic figure of characteristic fluctuations due to 
one faint object transiting across the other 
in front of their host star. 
}
\label{schematic}
\end{figure}

\begin{figure}
    \FigureFile(120mm,100mm){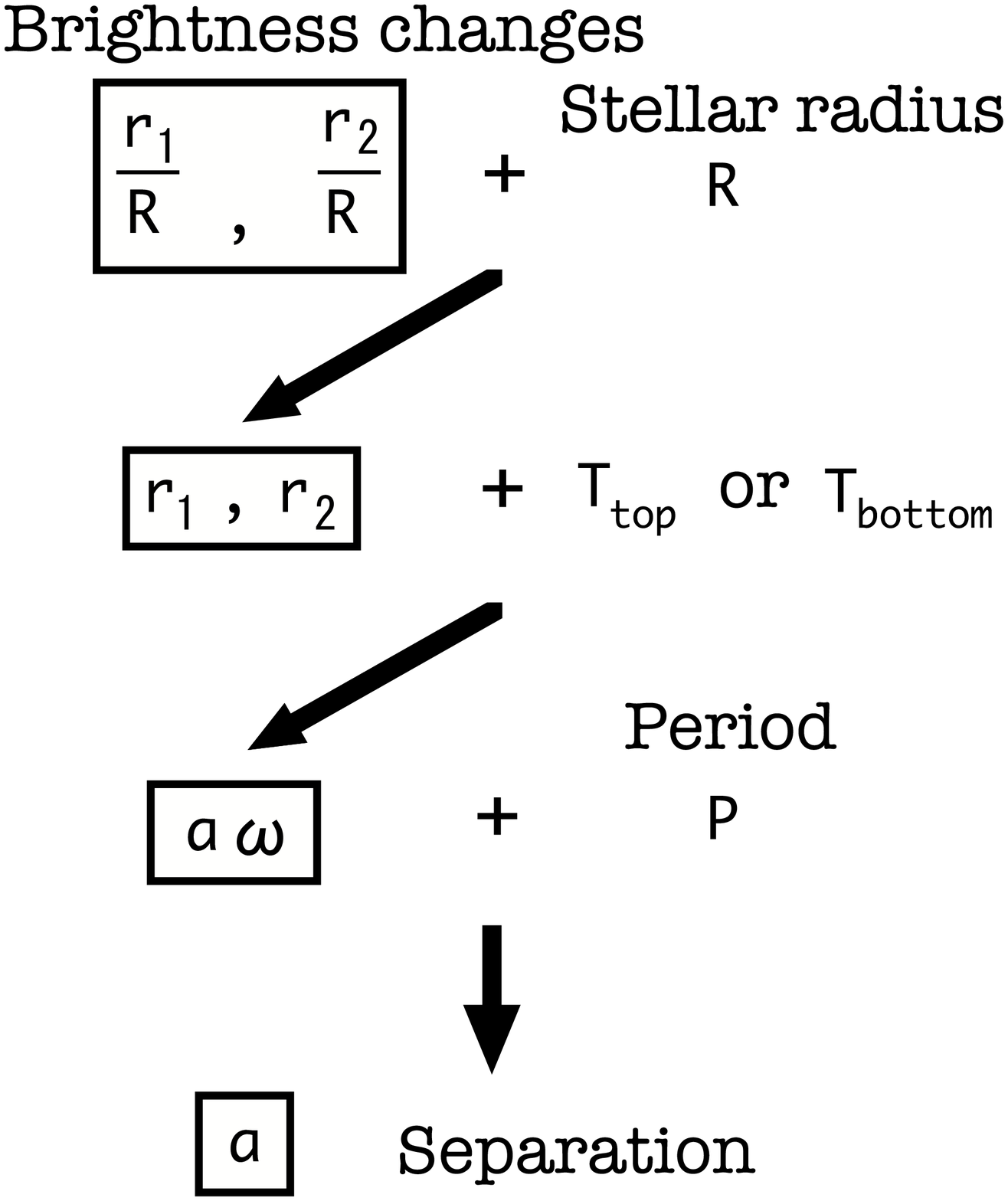}
\caption{ 
Flow chart of parameter determinations.  
Starting from measurements of brightness changes, 
the separation $a$ is eventually determined 
for a fast spin case. 
}
\label{flow-chart}
\end{figure}

Before closing this subsection, we briefly mention 
the time scale of the brightness fluctuation. 
The full width of a ``hill'' $T_{hill}$ 
corresponds to the crossing time 
of two planets as $2r/a\omega$. 
We thus obtain 
\begin{equation}
a\omega = \frac{2 r}{T_{hill}} . 
\label{aomega}
\end{equation}
By measuring the width, therefore, 
$a\omega$ can be determined directly and independently 
only for the fast spin case that produces spiky patterns. 
To be more precise, 
the full width of a ``hill'' at top and bottom 
are expressed as (See also Figure $\ref{schematic}$) 
\begin{eqnarray}
T_{top} &=& \frac{2(r_1-r_2)}{a\omega}, 
\label{Ttop}
\\
T_{bottom} &=& \frac{2(r_1+r_2)}{a\omega} . 
\label{Tbottom}
\end{eqnarray}
Only for symmetric binaries ($r_1=r_2$), 
we have $T_{top}=0$ and thus true spikes. 
Otherwise, truncated spikes (or ``hills'') appear. 
With $r_1$ and $r_2$ determined from brightness changes, 
measuring either $T_{top}$ or $T_{bottom}$ provides $a\omega$. 
This can be verified in Figure $\ref{lightcurve-3}$.
Figure $\ref{flow-chart}$ shows a flow chart of 
the parameter determinations that are discussed above. 

The half width for giant planets is about 
\begin{equation}
\frac{r}{a\omega} 
\sim 
5 \times 10^3 
\left(
\frac{r}{5 \times 10^4 \mbox{km}}\frac{10 \mbox{km/s}}{a\omega} 
\right) 
\mbox{sec.} 
\end{equation}
Therefore, detections of such fluctuations  
due to mutual transits of extrasolar binary planets 
require frequent observations, 
say every hour. 
Furthermore, more frequency (e.g., every ten minutes) 
is necessary for parameter estimations of the binary. 

\begin{figure}
    \FigureFile(130mm,120mm){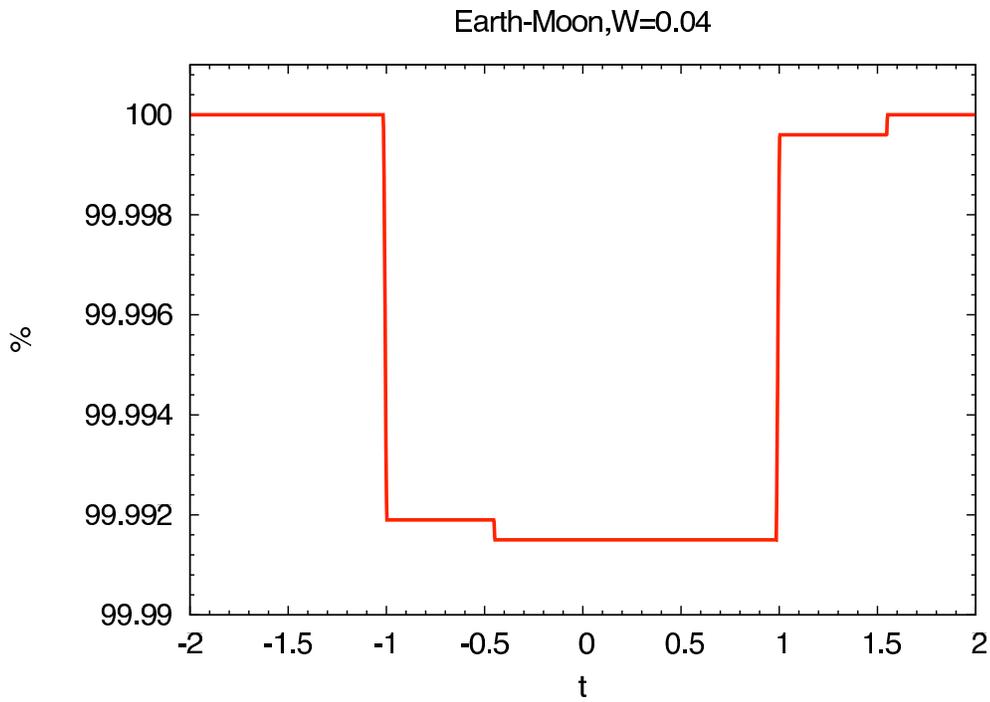}
\caption{
Light curve by transits of a Earth-Moon type system. 
The parameters are all assumed to be 
the same as those for the Earth and Moon in our solar system. 
Hence we have $r_1/R \sim 9 \times 10^{-3}$, 
$r_2 \sim 3 \times 10^{-3}$ and $W=0.04$ in this figure. 
}
\label{lightcurve-EM}
\end{figure}

Let us mention a connection of the present result 
with current space telescopes. 
Decrease in apparent luminosity due to the secondary planet 
is $O(r_2^2/R^2)$. 
Besides the time resolution (or observation frequency) and 
mission lifetimes, 
detection limits by {\it COROT} 
with the achieved accuracy 
of photometric measurements (700 ppm in one hour) 
could put $r_2/R \sim 2 \times 10^{-2}$. 
The nominal integration time is 32 sec. but 
co-added over 8.5 min. except for 1000 selected targets 
for which the nominal sampling is preserved. 
By the {\it Kepler} mission 
with expected 10 ppm differential sensitivity 
for solar-like stars with $m_V=12$, 
the lower limit will be reduced to $r_2/R \sim 3 \times 10^{-3}$. 
An analogy of the Earth-Moon 
($r_2/R \sim 2.5 \times 10^{-3}$, $W \sim 0.03$)
and 
Jupiter-Ganymede ($r_2/R \sim 4 \times 10^{-3}$, $W \sim 0.8$) 
will be marginally detectable. 
Figure $\ref{lightcurve-EM}$ shows a light curve due to 
an analogy of the Earth-Moon system. 
Observations both with high frequency 
(at least during the time of transits) 
and with good photometric sensitivity are desired 
for future detections of mutual transits.  
{\it COROT} satisfies these requirements and 
thus has a chance to find mutual transits. 
{\it Kepler} (with CCDs readout every three seconds) 
is one of the most suitable missions to date for the goal.

\subsection{Event rate and possible bounds on 
Earth-Moon-like systems} 
The probability of detecting mutual transits 
is expressed as $p=p_1 p_2 p_3$, 
where  
the probability for an object (with orbital period $P_{CM}$) 
transiting its host star during the observed time $T_{obs}$ 
is denoted as $p_1 = T_{obs}/P_{CM}$, 
that for one component transiting (or occulting) 
the other during the eclipse with duration $T_E$ 
is denoted as $p_2 = T_E/P$ for slow cases 
($p_2 = 1$ for $P < T_E$), 
and 
that for a condition that an observer 
is located in directions where the eclipse 
can be seen  
is denoted as $p_3 = \theta_{max}/(\pi/2)$.  
Here, the maximum angle from the orbital plane becomes 
$\theta_{max} \equiv R/a_{CM}$. 
Hence we obtain 
\begin{equation}
p =  
\frac{2 R T_E T_{obs}}{\pi a_{CM} P P_{CM}} , 
\label{p}
\end{equation}
which becomes 
$p \sim 6 \times 10^{-5} T_{obs} \mbox{yr}^{-1}$ 
for an Earth-Moon-like system. 
We thus need to monitor a number of stars ($N_S > 10^4$). 
Let $f$ denote the fraction of such systems 
for an averaged star. 
We have the expected events $n$ for observing $N_S$ stars 
during $T_{obs}$ as $n=f p N_S$. 
Therefore, 
monitoring $10^5$ stars for three years with {\it Kepler} 
may lead to the discovery of a second Earth-Moon-like system 
if the fraction is larger than 0.05, 
or 
it may put upper limits on the fraction as 
$f < (p N_S)^{-1} \sim 0.05 (3 \mbox{yr}/T_{obs}) (10^5/N_S)$.  

We should note that there exist constraints due to some physical
mechanisms on our parameters, especially the orbital separation. 
For instance, the companion's orbital radius must be 
larger than the Roche limit and 
smaller than the Hill radius 
(e.g., Danby 1988, Murray and Dermott 2000). 
These stability conditions are satisfied by the Earth-Moon system. 
Therefore, we can use Eq. ($\ref{p}$) for Earth-Moon analogies. 
For general cases such as ``exoearth-exomoon'' systems 
that have much smaller separations or are located at 
much shorter distance from their parent star, 
however, we have to take account of corrections due to 
certain physical constraints 
(e.g., Sartoretti and Schneider 1999 for estimates 
of such conditional probabilities with incorporating 
the Roche and Hill radii).

\section{Conclusion} 
We have shown that light curves by 
mutual transits of extrasolar planets depend strongly on 
a planet-companion's spin velocity, and 
especially for small separation cases where 
occultation of one faint object by the other 
transiting a parent star causes an apparent increase 
in light curves 
and characteristic fluctuations appear. 
We have shown also that extrasolar mutual transits provide 
a complementary method for measuring the radii 
of two transiting objects, 
their separation and mass,  
and consequently for identifying them as a true binary, 
planet-satellite system or others. 
Event rates and possible bounds on the fraction of Earth-Moon-like 
systems have been presented. 
Up to this point, we have considered only the obscuration effect 
in a simplistic manner. 
When actual light curves are analyzed, we should incorporate 
(1) a small deviation of the inclination angle from 90 degrees,  
(2) elliptical motions of the binary 
and 
(3) perturbations as three (or more)-body interactions 
(e.g., Danby 1988, Murray and Dermott 2000). 
Limb darkenings also should be taken into account.



\clearpage
\begin{table}
\caption{
List of quantities characterizing a system in this paper. 
}
\label{table1}
  \begin{center}
    \begin{tabular}{ll}
Symbol & Definition \\
\hline
$P_{CM}$ & Orbital period around a host star \\
\hline 
$P$ & Spin period of a planet-companion system\\
\hline
$\omega$ & Angular velocity of a planet-companion system 
$(=2\pi/P)$ \\
\hline 
$a_{CM}$ & Distance of planet-companion's center of mass from 
their host star \\
\hline 
$a$ & Separation of a planet-companion system\\
\hline 
$a_{\perp}$ & Apparent separation of a planet-companion system\\
\hline 
$R$ & Host star's radius \\
\hline 
$r_1$ & Planet's radius\\
\hline 
$r_2$ & Companion's radius\\
\hline 
$m_1$ & Planet's mass\\
\hline 
$m_2$ & Companion's mass\\
\hline 
$m_{tot}$ & $m_1 + m_2$\\
\hline 
$x_{CM}$ & Transverse position of a planet-companion's center of mass\\
\hline 
$x_1$ & Planet's transverse position\\
\hline 
$x_2$ & Companion's transverse position\\
\hline 
$t_0$ & Time at the maximum apparent separation of planet-companion\\
\hline 
$T_{top}$ & Time duration: width of a hill's top in light curves\\
\hline 
$T_{bottom}$ & Time duration: width of a hill's bottom in light curves\\ 
\hline 
$T_{12}$ & Time lag between the first and second transits\\ 
\hline 
$p$ & Detection probability for a given set of parameters\\
\hline
$f$ & Fraction of Earth-Moon-like systems for an averaged star \\
    \end{tabular}
  \end{center}
\end{table}

\clearpage

 \clearpage

\clearpage

\clearpage

\clearpage

\end{document}